\documentclass{aa}
\usepackage{txfonts}
\usepackage{graphicx}
\usepackage{natbib}
\usepackage[figuresright]{rotating}
\bibpunct{(}{)}{;}{a}{}{,}
\begin{document}

\title{The age of the Galactic thin disk from Th/Eu
nucleocosmochronology}

\subtitle{III. Extended sample\thanks{Based on observations
collected at the European Southern Observatory, La Silla, Chile,
under the ESO programs and the ESO-Observat\'orio Nacional,
Brazil, agreement, and at the Observat\'orio do Pico dos Dias,
operated by the Laborat\'orio Nacional de Astrof\'{\i}sica/MCT,
Brazil.}\fnmsep\thanks{Full Table~\ref{tab:sample_ews} is only
available in electronic form at the CDS via anonymous ftp to {\tt
cdsarc.u-strasbg.fr (130.79.128.5)} or via {\tt
http://cdsweb.u-strasbg.fr/cgi-bin/qcat?}}}

\titlerunning{The age of the Galactic thin disk from Th/Eu
nucleocosmochronology. III}

\author{E.F. del Peloso\inst{1}\fnmsep\inst{2} \and L. da Silva\inst{1} \and
G.F. Porto de Mello\inst{2} \and L.I. Arany-Prado\inst{2}}

\offprints{E.F. del Peloso}

\institute{Observat\'orio Nacional/MCT, Rua General Jos\'e
Cristino
77, 20921-400 Rio de Janeiro, Brazil\\
\email{epeloso@on.br, licio@on.br} \and Observat\'orio do
Valongo/UFRJ, Ladeira
do Pedro Ant\^onio 43, 20080-090 Rio de Janeiro, Brazil\\
\email{gustavo@ov.ufrj.br, lilia@ov.ufrj.br}}

\date{Received <date> / Accepted <date>}

\abstract{The first determination of the age of the Galactic thin
disk from Th/Eu nucleocosmochronology was accomplished by us in
Papers~I and II. The present work aimed at reducing the age
uncertainty by expanding the stellar sample with the inclusion of
seven new objects -- an increase by 37\%. A set of [Th/Eu]
abundance ratios was determined from spectral synthesis and merged
with the results from Paper~I. Abundances for the new, extended
sample were analyzed with the aid of a Galactic disk chemical
evolution (GDCE) model developed by us is Paper~II. The result was
averaged with an estimate obtained in Paper~II from a conjunction
of literature data and our GDCE model, providing our final,
adopted disk age $T_{\mathrm{G}}=(8.8\pm1.7)~\mbox{Gyr}$ with a
reduction of 0.1~Gyr (6\%) in the uncertainty. This value is
compatible with the most up-to-date white dwarf age determinations
($\lesssim10$~Gyr). Considering that the halo is currently
presumed to be $(13.5\pm0.7)~\mathrm{Gyr}$ old, our result prompts
groups developing Galactic formation models to include an hiatus
of $(4.7\pm1.8)~\mathrm{Gyr}$ between the formation of halo and
disk. }

\maketitle

\section{Introduction}

The age of the Galactic thin disk\footnote{All references to the
\emph{Galactic disk} must be regarded, in this work, as references
to the \emph{thin} disk, unless otherwise specified.} is an
important constraint for Galactic formation models. It is usually
estimated by dating the oldest open clusters with isochrones or
white dwarfs with cooling sequences. These methods are strongly
dependent on stellar evolution models and on numerous physical
parameters known at different levels of uncertainty.
Nucleocosmochronology is only weakly dependent on main sequence
stellar evolution models, allowing for a nearly independent
crosscheck of other techniques.

We were the first to determine an age for the Galactic disk from
Th/Eu nucleocosmochronology -- \citealt{delpelosoetal05a}
(Paper~I) and \citealt{delpelosoetal05b} (Paper~II), based on the
PhD thesis of one of us \citep{delpeloso03}. This work, the last
part in a series of three articles, aims at reducing the
uncertainty in this determination by expanding the stellar sample
from Papers~I and II with objects observed only after the
publication of \citet{delpeloso03} and deriving a new age from
this extended sample. With this intent, we determined [Th/Eu]
abundance ratios for a sample of Galactic disk stars and employed
Galactic disk chemical evolution models developed by us in the
chronological analysis.\footnote{In this paper we obey the
customary spectroscopic notation: abundance ratio
$\mbox{[A/B]}\equiv\log_{10}(N_{\mathrm{A}}/N_{\mathrm{B}})_{\mathrm{star}}
-\log_{10}(N_{\mathrm{A}}/N_{\mathrm{B}})_{\mbox{\scriptsize\sun}}$,
where $N_{\mathrm{A}}$ and $N_{\mathrm{B}}$ are the abundances of
elements A and B, respectively, in atoms~cm$^{-3}$.}

\section{Sample selection, observations and data reduction}

Sample selection, observations and data reduction were carried out
following exactly the same procedures described in detail in
Paper~I. In what follows we provide a brief overview of these
topics.

\begin{figure}
\resizebox{\hsize}{!}{\includegraphics*{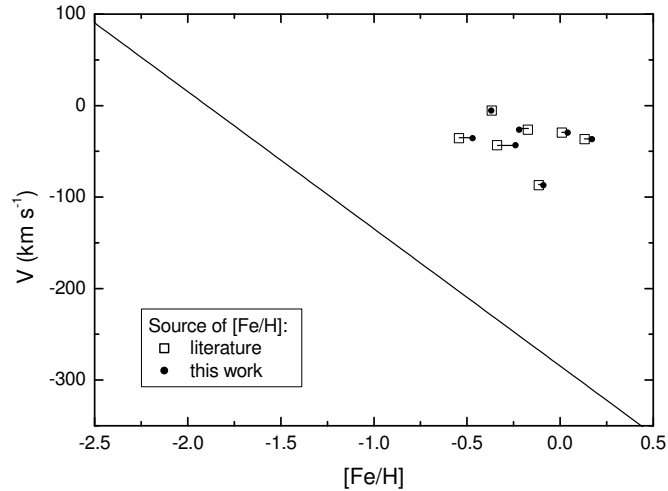}} \caption{V
vs. [Fe/H] diagram for the sample stars. The diagonal line is a
cut off between the halo and the disk populations (below and
above, respectively). Velocity components V were calculated by us.
The metallicities initially employed, represented as open squares,
were taken from the literature (Table~\ref{tab:photometry}); small
filled circles represent metallicities derived in this work.}
\label{fig:kinematics}
\end{figure}

The stellar sample of this work is composed of seven F8--G5 dwarfs
and subgiants (Table~\ref{tab:sample}). As the objective of this
work is the determination of the age of the Galactic \emph{disk},
we performed a kinematic analysis of the objects to help ensure
that they do not belong to the Galactic halo. We calculated the
objects' $U$, $V$, and $W$ spatial velocity components
(Table~\ref{tab:kinematic_data}) and constructed a $V$ vs. [Fe/H]
diagram (Figure~\ref{fig:kinematics}) using metallicities from the
literature (Table~\ref{tab:photometry}). According to
\citet{schusteretal93}, objects located above the displayed
cut-off line belong to the Galactic halo, whereas those located
below the line belong to the halo. It can be seen that all sample
stars are located far above the cut-off line. For a star to cross
the line, the literature metallicities would have to have been
\emph{overestimated} by at least 1.2~dex, which is very unlikely.
After having determined our own metallicities
(Table~\ref{tab:adopted_atmospheric_parameters}), we confirmed
that results from the literature agree with our values to 0.1~dex.

\begin{table*} \caption[]{Selected stellar sample.}
\label{tab:sample}
\begin{tabular}{ l r r l c c r @{.} l r @{.} l c }
\hline \hline HD & HR & HIP & Name & R.A. & DEC &
\multicolumn{2}{c}{Parallax} & \multicolumn{2}{c}{V} &
Spectral type\\
 & & & & 2000.0 & 2000.0 & \multicolumn{2}{c}{(mas)} &\multicolumn{2}{c}{} &and\\
 & & & & (h m s) & (d m s) & \multicolumn{2}{c}{}&\multicolumn{2}{c}{}& luminosity class\\
\hline 1461 & 72 & 1499 &--& 00 18 42 & $-$08 03 11 &  42&67 & 6&46 & G5 V\\
157\,089&  -- &84\,905 &--& 17 21 07 &   +01 26 35 & 25&88 & 6&95 &G0--2 V\\
162\,396&6649 & 87\,523 & --& 17 52 53 & $-$41 59 48 & 30&55 &
6&20 &F8 IV--V\\
189\,567& 7644 & 98\,959 & --& 20 05 33 & $-$67 19 15 & 56&45 &
6&07 & G3 V\\
193\,307& 7766 &100\,412 & -- & 20 21 41 & $-$49 59 58 & 30&84 &
6&27 & G0 V\\
196\,755& 7896 &101\,916 &$\kappa$ Del& 20 39 08 & +10 05 10 & 33&27 & 5&05 &G5 IV\\
210\,918&8477&109\,821 &--& 22 14 39 & $-$41 22 54 & 45&19 & 6&23 & G5 V\\
\hline
\end{tabular}

{References: Coordinates: SIMBAD (FK5 system); parallaxes:
Hipparcos catalogue \citep{hipparcos}; visual magnitudes: Bright
Star Catalogue \citep{brightstar} for stars with an HR~number and
SIMBAD for those without; spectral types and luminosity classes:
Michigan Catalogue of HD~Stars
\citep{michigancatalog1,michigancatalog2,michigancatalog5} for all
stars, with the exception of HD~196\,755 (Bright Star Catalogue).}
\end{table*}

\begin{table}
\caption[]{Radial velocities ($RV$) and spatial velocity
components ($U$, $V$, and $W$) of the sample stars, in a
right-handed Galactic system and relative to the LSR. All values
are in km~s$^{-1}$.} \label{tab:kinematic_data}
\begin{tabular}{ l
r @{.} l r @{.} l r @{.} l r @{.} l } \hline \hline
HD & \multicolumn{2}{c}{$RV$} & \multicolumn{2}{c}{$U$} &  \multicolumn{2}{c}{$V$} & \multicolumn{2}{c}{$W$}\\
\hline
1461 &$-$9 & 3 & $-$15 & 3 $\pm$0.7 & $-$36 & 7 $\pm$1.0 &   +17 & 0 $\pm$0.4\\
157\,089& $-$161& 1 &$-$156& 0 $\pm$1.1 & $-$35 & 5 $\pm$1.0 &$-$2 & 8 $\pm$1.8\\
162\,396&$-$16& 4 &$-$13& 3 $\pm$0.4 & $-$5  & 5 $\pm$0.7 & $-$24& 0 $\pm$1.3\\
189\,567&$-$10& 0 & $-$59 & 7 $\pm$1.6 & $-$26 & 4 $\pm$0.7 &$-$42& 6 $\pm$2.0\\
193\,307& +18 & 6 &   +47 & 4 $\pm$1.2 & $-$43 & 3 $\pm$1.6 & +40 & 8 $\pm$2.0\\
196\,755&$-$52& 7 &$-$48& 0 $\pm$0.8 & $-$29 & 5 $\pm$0.6 & $-$11 & 2 $\pm$1.0\\
210\,918&$-$18 & 6 & $-$37 & 4 $\pm$0.9 & $-$86 & 9 $\pm$1.6 & $-$1  & 8 $\pm$0.7\\
\hline
\end{tabular}
\end{table}

\begin{table*}
\caption[]{Photometric indices and metallicities for all sample
stars. Metallicities were taken from the literature for use in the
kinematic characterization of the sample and as input for the
first step in the iterative determination of atmospheric
parameters.} \label{tab:photometry}
\begin{tabular}{l @{\hspace{2em}} c c @{\hspace{2em}} c c @{\hspace{2em}}
c c @{\hspace{2em}} c c @{\hspace{2em}} c c @{\hspace{2em}} c c}
\hline \hline HD & ($B-V$) & Ref. & ($b-y$) & Ref. & $\beta$ &
Ref. & ($V-K$) &
Ref. & ($B_{\mathrm{T}}-V_{\mathrm{T}}$)& Ref. & $[$Fe/H$]$ & Ref.\\
\hline
1461   & 0.68 & 1 & 0.422 & 3 &2.596& 4 & 1.518 &  7 &0.764& 9 &   +0.13 & 10\\
157\,089& 0.57 & 2 & 0.379 & 3 & 2.584 &  5 & 1.434 &  7 & 0.619 & 9  &$-$0.54& 11\\
162\,396 & 0.54 & 1 & 0.347 & 4 &2.612& 4 & -- & --& 0.563 & 9  &$-$0.37& 10\\
189\,567 & 0.64 & 1 & 0.406 & 3 & 2.583 &  6 & 1.516 &  7 &0.718& 9 & $-$0.17 & 10\\
193\,307 & 0.55 & 1 & 0.365 & 3 &2.604& 5 &   --  & -- &0.617& 9 & $-$0.34 & 10\\
196\,755 & 0.72 & 1 & 0.432 & 3 &   --  & -- & --  & -- & 0.765 & 9  &   +0.01 & 10\\
210\,918 & 0.65 & 1 & 0.404 & 3 &2.590&  6&1.500& 8 &   --  & 9  & $-$0.12 & 10\\
 \hline
\end{tabular}

References: 1~-~\citet{brightstar}; 2~-~\citet{mermilliod87};
3~-~\citet{olsen94b};
4~-~\citet{gronbech&olsen76,gronbech&olsen77};
5~-~\citet{olsen83}; 6~-~\citet{schuster&nissen88};
7~-~\citet{dibenedetto98}; 8~-~\citet{koornneef83};
9~-~\citet{hipparcos}; 10~-~\citet{taylor03};
11~-~\citet{cayreldestrobeletal01}.
\end{table*}

High resolution, high signal-to-noise ratio spectra were obtained
for all objects with the Fiber-fed Extended Range Optical
Spectrograph \citep[FEROS;\ ][]{kauferetal99} fed by the 1.52~m
European Southern Observatory (ESO) telescope, in the
ESO-Observat\'orio Nacional, Brazil, agreement (March and August
2001). Spectra were also obtained with a coud\'e spectrograph fed
by the 1.60~m telescope of the Observat\'orio do Pico dos Dias
(OPD), LNA/MCT, Brazil (May and October 2000; May, August, and
October 2002), and with the Coud\'e \'Echelle Spectrometer (CES)
fiber-fed by ESO's 3.60~m telescope (August 2003). FEROS spectra
are reduced automatically during observation by a script executed
using the European Southern Observatory Munich Image Data Analysis
System (ESO-MIDAS) immediately after the CCD read out. OPD and CES
spectra were reduced by us using the Image Reduction and Analysis
Facility (IRAF\footnote{IRAF is distributed by the National
Optical Astronomy Observatories, which are operated by the
Association of Universities for Research in Astronomy, Inc., under
cooperative agreement with the National Science Foundation.}),
following the usual steps of bias, scattered light and flat field
corrections, and extraction.

\section{Atmospheric parameters}

A set of homogeneous, self-consistent atmospheric parameters was
determined for the sample stars, following the procedure described
in detail in Paper~I. Effective temperatures were derived from
photometric calibrations (Table~\ref{tab:photometry}) and
H$\alpha$ profile fitting; surface gravities were obtained from
$T_{\mathrm{eff}}$, stellar masses and luminosities;
microturbulence velocities and metallicities were obtained from
detailed, differential spectroscopic analysis, relative to the
Sun, using EWs of \ion{Fe}{i} and \ion{Fe}{ii} lines. The final,
adopted values of all atmospheric parameters are presented in
Table~\ref{tab:adopted_atmospheric_parameters}.

\begin{table}
\caption[]{Adopted atmospheric parameters, including the
photometric and H$\alpha$ effective temperatures used to obtain
the adopted mean values, and the stellar masses used to obtain the
surface gravities.} \label{tab:adopted_atmospheric_parameters}
\begin{tabular}{ l @{\hspace{0.70em}} c @{\hspace{0.70em}} c @{\hspace{0.53em}}
c @{\hspace{0.53em}} c @{\hspace{0.70em}} c @{\hspace{0.70em}} c
@{\hspace{0.70em}} c @{\hspace{0.70em}} c @{\hspace{0.70em}} c }
\hline \hline HD & \multicolumn{3}{c}{$T_{\mathrm{eff}}$ (K)} & &
$m_{\mbox{\scriptsize\sun}}$ & $\log g$
& [Fe/H] & $\xi$ (km~s$^{-1}$)\\
\cline{2-4} & Phot. & H$\alpha$ & MEAN & &  & & & \\
 \hline
1461 & 5727 & 5705 & 5717 & & 1.01 & 4.33 &   +0.17 & 1.20  \\
157\,089& 5827 &5742& 5785 && 0.93 & 4.09 &$-$0.47 & 0.90 \\
162\,396& 6072 & 5979 & 6026 && 1.04 & 4.08 & $-$0.37 & 1.36 \\
189\,567& 5704 & 5715 & 5710 && 0.90 & 4.36 & $-$0.22 & 1.01 \\
193\,307& 5999 & 5956 & 5978 && 1.05 & 4.11 & $-$0.24 &  1.25 \\
196\,755& 5665 & 5613 & 5639 && 1.58 & 3.70 & +0.04 & 1.42 \\
210\,918& 5733 & 5708 & 5721 && 0.95 & 4.27 & $-$0.09 & 1.15 \\
 \hline
\end{tabular}
\end{table}

\section{Abundances of contaminating elements}
\label{sec:contamin_abund}

Eu and Th abundances were determined by spectral synthesis of the
\ion{Eu}{ii} line at 4129.72~\AA\ and of the \ion{Th}{ii} line at
4019.13~\AA, respectively. In the synthesis calculations,
abundances of the elements that contaminate these spectral regions
(Ti, V, Cr, Mn, Co, Ni, Ce, Nd, and Sm) were kept fixed in the
values determined using EWs and the atmospheric parameters
obtained by us; see Paper~I for a full description of the employed
method. Table~\ref{tab:sample_ews} presents a sample of the EW
data. Its complete content, composed of the EWs of all measured
lines, for the Sun and all sample stars, is only available in
electronic form at the CDS. Column~1 lists the central wavelength
(in angstroms), Col.~2 gives the element symbol and degree of
ionization, Col.~3 gives the excitation potential of the lower
level of the electronic transition (in eV), Col.~4 presents the
solar $\log gf$ derived by us, and the subsequent columns present
the EWs, in m\AA, for the Sun and the other stars, from HD~1461 to
HD~210\,918.

\begin{table}
\caption[]{A sample of the EW data. The complete content of this
table is only available in electronic form at the CDS. For a
description of the columns, see text
(Sect.~\ref{sec:contamin_abund}).} \label{tab:sample_ews}
\begin{tabular}{@{} c c c c c @{\hspace{1em}} c @{\hspace{1em}}
c @{}} \hline \hline  $\lambda$~(\AA) & Element &
$\chi$~(eV) & $\log gf$ &
Sun & $\cdots$ & HD~210\,918\\
 \hline
5668.362 & \ \ion{V}{i} & 1.08 & $-$0.920 & \ \ 8.5 & $\cdots$ & 11.9\\
5670.851 & \ \ion{V}{i} & 1.08 & $-$0.452 & 21.6 & $\cdots$ & 23.7\\
5727.661 & \ \ion{V}{i} & 1.08 & $-$0.657 & \ \ 9.5 & $\cdots$ & 12.1\\
$\vdots$ & $\vdots$ & $\vdots$ & $\vdots$ & $\vdots$ & $\vdots$ & $\vdots$ \\
5427.826 & \ion{Fe}{ii} & 6.72 & $-$1.371 & \ \ 6.4 & $\cdots$ & \ \ 0.0\\
6149.249 & \ion{Fe}{ii} & 3.89 & $-$2.711 & 40.9 & $\cdots$ & 39.5\\
\hline
\end{tabular}
\end{table}

\begin{figure*}
\centering
\includegraphics*[width=17cm]{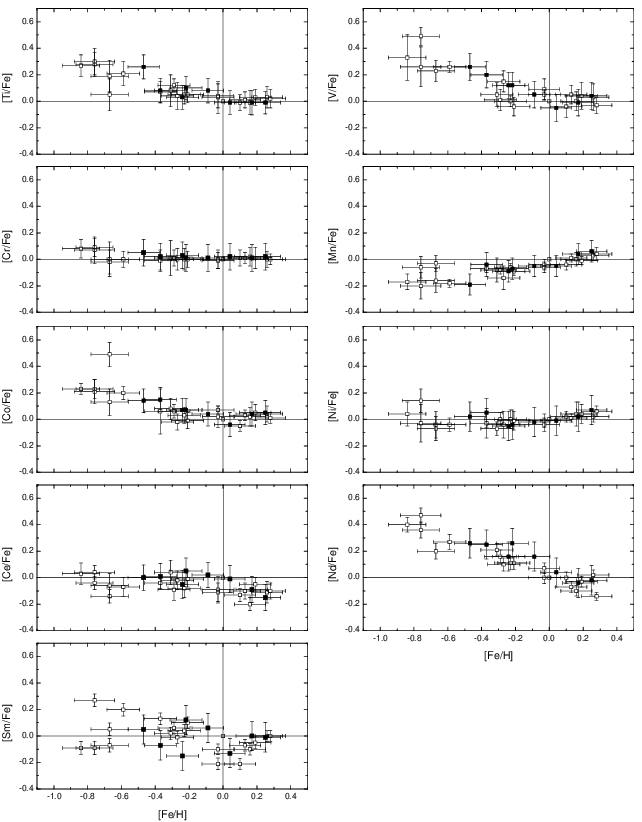}
\caption{Abundance patterns for all contaminating elements. Filled
squares are data for this work; open squares are data from
Paper~I.} \label{fig:abundances}
\end{figure*}

Abundance results are presented in Table~\ref{tab:abundances}. No
detailed uncertainty assessment was carried out. We have rather
adopted an average of the uncertainties presented in Paper~I:
0.08~dex for Mn, 0.09~dex for Fe, Ti, and Co, 0.10~dex for V, Cr,
Ce, and Nd, and 0.11~dex for Ni and Sm. These values are used as
error bars in Fig.~\ref{fig:abundances}, which shows the abundance
patterns for all elements. Note that the abundances for the sample
of this work (filled squares) agree very well with those from
Paper~I (open squares).

\begin{table*}\caption[]{Fe, Ti, V, Cr, Mn, Co, Ni, Ce, Nd, and Sm abundances, relative to H. N is
the number of absorption lines effectively used for each abundance
determination.} \label{tab:abundances}
\begin{tabular}{ l @{\hspace{1em}}
c @{\hspace{1em}} c @{\hspace{1em}} c @{\hspace{1em}} c
@{\hspace{1em}} c @{\hspace{1em}} c @{\hspace{1em}} c
@{\hspace{1em}} c @{\hspace{1em}} c @{\hspace{1em}} c
@{\hspace{1em}} c @{\hspace{1em}} c @{\hspace{1em}} c
@{\hspace{1em}} c } \hline \hline HD & [\ion{Fe}{i}/H] & N &
[\ion{Fe}{ii}/H] & N & [Fe/H] & N &  [Ti/H] & N & [V/H] & N &
[Cr/H] & N & [Mn/H] & N\\
\hline 1461 & +0.16 & 52 &+0.19 & 8 & +0.17 & 60 &
+0.16 & 28 & +0.16 & 5 & +0.18 & 19 & +0.21 & 6\\
157\,089 & $-$0.47 & 31 & $-$0.47 & 7 & $-$0.47 & 38 & $-$0.21 &
20 & $-$0.21 & 3 & $-$0.42 & 18 & $-$0.66 & 4\\
162\,396& $-$0.38 & 41 & $-$0.33 & 8 & $-$0.37 & 49 & $-$0.29 & 18
& $-$0.17 & 2 & $-$0.35 & 16 & $-$0.41 & 5\\
189\,567& $-$0.23 & 44 & $-$0.20 & 7 & $-$0.22 & 51 & $-$0.12 & 22
& $-$0.10 & 3 & $-$0.21 & 19 & $-$0.29 & 6\\
193\,307& $-$0.24 & 45 & $-$0.22 & 9 & $-$0.24 & 54 &
$-$0.21 & 19 & $-$0.12 & 4 & $-$0.21 & 19 & $-$0.33 & 6\\
196\,755& +0.04 & 45 & +0.06 & 9 & +0.04 & 54 & +0.03 & 28
& $-$0.01 & 5 & +0.06 & 20 & $-$0.01 & 6\\
210\,918& $-$0.09 & 47 & $-$0.09 & 7 & $-$0.09 & 54 & $-$0.01 & 28
& $-$0.04 & 5 & $-$0.08 & 18 & $-$0.14 & 6\\
\hline
\end{tabular}

\vspace{0.3cm}

\begin{tabular} { l @{\hspace{1em}} c @{\hspace{1em}} c @{\hspace{1em}} c
@{\hspace{1em}} c @{\hspace{1em}} c @{\hspace{1em}} c
@{\hspace{1em}} r @{\hspace{1em}} c @{\hspace{1em}} c
@{\hspace{1em}} c } \hline\hline HD & [Co/H] & N & [Ni/H] & N &
[Ce/H] & N & [Nd/H] & N & [Sm/H] & N\\
\hline 1461 & +0.21 & 8 & +0.19 & 10 & +0.08 & 5 & +0.13 & 2
& +0.17 & 1\\
157\,089& $-$0.33 & 7 & $-$0.45 & 7 & $-$0.47 & 5 & $-$0.21 & 2 &
$-$0.42 & 1\\
162\,396& $-$0.22 & 3 & $-$0.32 & 7 & $-$0.36 & 4 & $-$0.12 & 2 &
$-$0.44 & 1\\
189\,567& $-$0.15 & 8 & $-$0.26 & 9 & $-$0.17 & 4 & +0.04 & 2 &
$-$0.10 & 1\\
193\,307& $-$0.17 & 6 & $-$0.29 & 8 & $-$0.29 & 5 & $-$0.08 & 2 &
$-$0.39 & 1\\
196\,755& +0.00 & 8 & +0.03 & 10 & +0.03 & 5 & +0.08 & 2 &
$-$0.09 & 1\\
210\,918& $-$0.05 & 8 & $-$0.11 & 9 & $-$0.07 & 5 & +0.07 & 2 &
$-$0.03 & 1\\
\hline
\end{tabular}
\end{table*}

\section{Eu and Th abundances}

Our determination of Eu abundances by spectral synthesis used
FEROS spectra. As the abundances used for age determination in
Paper~I were obtained with the CES coupled to the Coud\'e
\'Echelle Spectrograph (CAT), we converted our results to the CAT
system, using the equation derived in Sect.~5.1.1 of Paper~I (see
their Fig.~15):
$\mathrm{[Eu/H]}_{\mathrm{CAT+CES}}=-0.00462+0.96480\,\mathrm{[Eu/H]}_{\mathrm{FEROS}}$.
Table~\ref{tab:th_eu_abundances} presents the [Eu/H], [Th/H], and
[Th/Eu] abundance ratios for our sample. As uncertainty, we
adopted an average of the values presented in Paper~I: 0.04~dex
for [Eu/H], 0.11~dex for [Th/H], and 0.08~dex for [Th/Eu].

\begin{table}
\caption[]{[Eu/H], [Th/H], and [Th/Eu] abundance ratios.}
\label{tab:th_eu_abundances}
\begin{tabular} { l c c c }
\hline\hline HD & [Eu/H] & [Th/H] & [Th/Eu]\\
\hline
1461 & +0.09 & +0.08 & $-$0.01\\
157\,089 & $-$0.12 & $-$0.28 & $-$0.16\\
162\,396 & $-$0.23 & $-$0.12 & +0.11\\
189\,567 & +0.08 & +0.02 & $-$0.06\\
193\,307 & $-$0.13 & $-$0.01 & +0.12\\
196\,755 & +0.02 & +0.09 & +0.07\\
210\,918 & +0.03 & $-$0.01 & $-$0.04\\
\hline
\end{tabular}
\end{table}

\begin{figure}
\resizebox{\hsize}{!}{\includegraphics*{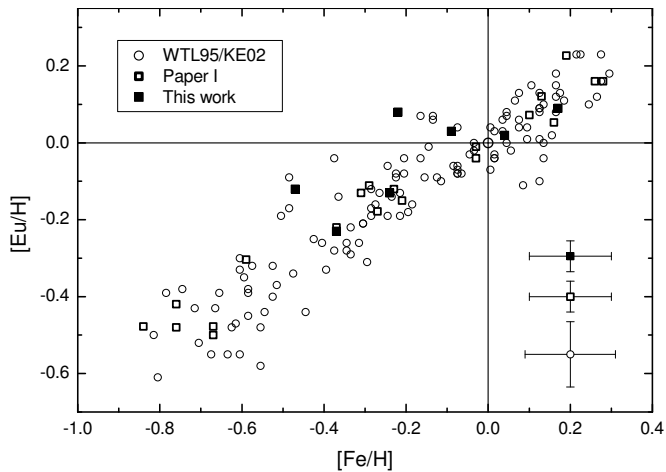}}
\caption{[Eu/H] vs. [Fe/H] diagram for our sample stars, and those
from Paper~I, WTL95, and KE02. Average error bars for the three
data sets are provided in the lower right corner.}
\label{fig:eu_h_fe_h_wtl95_ke02}
\end{figure}

In Fig.~\ref{fig:eu_h_fe_h_wtl95_ke02}, we compare our [Eu/H]
abundance ratios to those from \citet[\space WTL95]{woolfetal95},
\citet[\space KE02]{koch&edvardsson02}, and Paper~I. These are the
best works available with Eu abundances for Galactic disk stars,
in terms of care of analysis and sample size. Although two of our
objects (HD~1461 and HD~157\,089) exhibit considerable discrepancy
with Paper~I, our results present behavior and dispersion similar
to WTL95/KE02.

Th spectral synthesis employed CES spectra fed by the 3.60~m ESO
telescope, whereas the abundances in Paper~I were obtained with
the CES fed by the CAT. Our results were converted to the CAT
system, using the equation derived in Sects.~5.2.1 of Paper~I (see
their Fig.~20):
$\mathrm{[Th/H]}_{\mathrm{CAT+CES}}=-0.01500+0.86000\,\mathrm{[Th/H]}_{\mathrm{CAT+3.60~m}}$.

\begin{figure}
\resizebox{\hsize}{!}{\includegraphics*{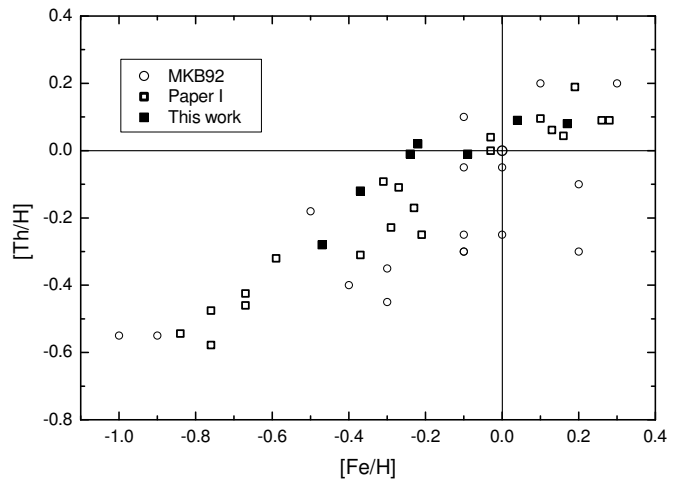}}
\caption{[Th/H] vs. [Fe/H] diagram for our sample stars, and those
from Paper~I and MKB92. Our results are in good accord with MKB92,
but show a lower dispersion.} \label{fig:th_h_fe_h_mkb92}
\end{figure}

In Fig.~\ref{fig:th_h_fe_h_mkb92}, we compare our [Eu/H] abundance
ratios to those from \citet[\space MKB92]{morelletal92} and
Paper~I. These are the best works available with Th abundances for
Galactic disk stars, in terms of care of analysis and sample size.
Our results are in good accord with them.

\section{Chronological analysis}

In order to estimate the age of the Galactic disk, we compared the
stellar [Th/Eu] abundance ratios with curves calculated using a
Galactic disk chemical evolution (GDCE) model. In this model,
developed by us based on \citet{pagel&tautvaisiene95}, it was
assumed that the so-called ``universality of the r-process
abundances'' is valid for second and third r-process peaks and can
be extended to the actinides. Such extension may not be
legitimate, as two ultra-metal-poor stars -- CS~31\,082-001
(\citealt{cayreletal01} and \citealt{hilletal02}) and
CS~30\,306-132 \citep{hondaetal04} -- have been recently shown to
have Th/Eu abundance ratios much higher than expected for their
age. This could indicate that they have been formed from matter
enriched in actinides, relatively to second r-process peak
elements. However, it is not yet clear if CS~31\,082-001 and
CS~30\,306-132 are merely chemically peculiar objects or if their
discrepancies could be present in other yet unobserved stars. A
detailed description of the GDCE model was presented in Paper~II.

\begin{figure}
\resizebox{\hsize}{!}{\includegraphics*{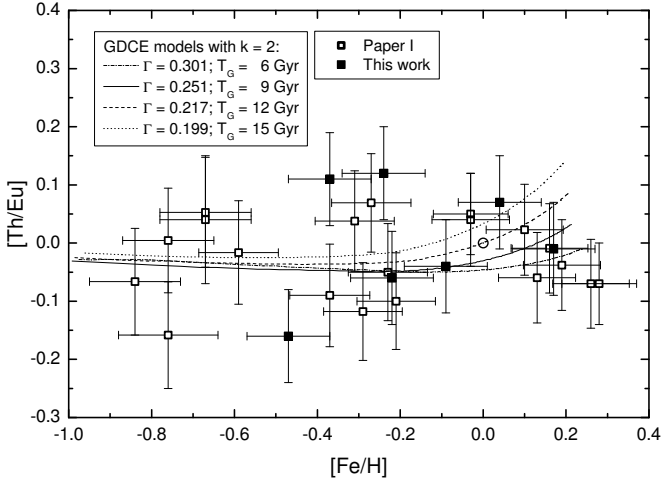}}
\caption{[Th/Eu] vs. [Fe/H] diagram for our sample stars and those
from Paper~I. Curves were calculated using our GDCE model for the
four different Galactic disk ages shown in the legend.}
\label{fig:th_eu_fe_h_curves}
\end{figure}

The abundances for our sample were determined in exactly the same
way as those from Paper~I, as the objective of this work is to
expand the sample used for chronological analysis. Accordingly, we
merged our abundance data with those from Paper~I, resulting in a
set of 28~objects. The theoretical model curves and the observed
abundance data are presented in Fig.~\ref{fig:th_eu_fe_h_curves}.

The age of the curve that best fits the observed data was computed
by minimizing the total deviation between the curves and the data.
26 of the 28~objects in the merged sample were actually used in
the determination; two objects from Paper~I were disregarded
because they are too metal-rich, falling out of the interval where
the curves are defined. Considering that in Paper~II 19~objects
were actually employed in the analysis, this work accomplishes a
37\%~increase in sample size. An uncertainty for the age, related
solely to the uncertainties in the abundances, was computed
through a Monte Carlo simulation. It must be noted that this is
only an assessment of the \emph{internal} uncertainty of the
analysis, and does not take into consideration the uncertainties
of the GDCE model itself, which are very difficult to estimate.
The uncertainty related to the model could very well be the main
source of age uncertainty. These procedures are fully described in
Paper~II. The final value obtained using the merged abundance data
set is $(8.8\pm1.8)~\mathrm{Gyr}$. This result, 0.6~Gyr larger and
with an uncertainty 0.1~Gyr lower than that of Paper~II, matches
very well the estimate obtained from literature data and the GDCE
model ($8.7^{+5.8}_{-4.1}~\mathrm{Gyr}$). These estimates were
combined using the maximum likelihood method, assuming that each
one follows a Gaussian probability distribution, which results in
a weighted average using the reciprocal of the square
uncertainties as weights. The final, adopted Galactic disk age is
$$T_{\mathrm{G}}=(8.8\pm1.7)~\mbox{Gyr.}$$

\section{Conclusions}

We determined [Th/Eu] abundance ratios for a sample of seven
Galactic disk F8--G5 dwarfs and subgiants. The analysis was
carried out in exactly the same way as that of Paper~I, so that we
could merge the data, resulting in a totally homogeneous extended
sample of 28~objects; 26 of these were actually used in the
nucleocosmochronological analysis. A GDCE model, developed in
Paper~II, was used in conjunction with the stellar abundance data
to compute an age for the Galactic disk:
$(8.8\pm1.8)~\mathrm{Gyr}$. In Paper~II, Th/Eu production and
solar abundance ratio data taken from the literature were analyzed
with our GDCE model, yielding $8.7^{+5.8}_{-4.1}~\mathrm{Gyr}$.
These two results were combined using the maximum likelihood
method, resulting in
$\mbox{FINAL\space}T_{\mathrm{G}}=(8.8\pm1.7)~\mbox{Gyr}$.

The inclusion of seven more stars in the abundance data base had
two main consequences: it increased the age obtained from our
stellar data by 0.6~Gyr, rendering it more compatible with the age
determined from literature data, and it decreased the uncertainty
of the final, adopted age by 0.1~Gyr, i.e., a 6\%~reduction. Our
result remains compatible with the most up-to-date white dwarf
ages derived from cooling sequence calculations, which indicate a
low age ($\lesssim10~\mathrm{Gyr}$) for the disk
\citep{oswaltetal95,bergeronetal97,leggettetal98,knoxetal99,hansenetal02}.
Considering that the age of the oldest halo globular clusters are
currently estimated at $(13.5\pm0.7)~\mathrm{Gyr}$
\citep{pontetal98,jimenez99,grattonetal03,krauss&chaboyer03}, an
hiatus of $(4.7\pm1.8)~\mathrm{Gyr}$ between the formation of halo
and disk must be taken into consideration in future Galactic
formation models.

\begin{acknowledgements}
The authors wish to thank the staff of the Observat\'orio do Pico
dos Dias, LNA/MCT, Brazil and of the European Southern
Observatory, La Silla, Chile. We thank R. de la Reza for his
contributions to this work. The suggestions of Dr.~N. Christlieb,
the referee, were greatly appreciated. EFP acknowledges financial
support from CAPES/PROAP, FAPERJ/FP (grant E-26/150.567/2003), and
CNPq/DTI (grant 382814/2004-5). LS thanks the CNPq, Brazilian
Agency, for the financial support 453529.0.1 and for the grants
301376/86-7 and 304134-2003.1. GFPM acknowledges financial support
from CNPq/Conte\'udos Digitais, CNPq/Institutos do
Mil\^enio/MEGALIT, FINEP/PRONEX (grant 41.96.0908.00),
FAPESP/Tem\'aticos (grant 00/06769-4), and FAPERJ/APQ1 (grant
E-26/170.687/2004).
\end{acknowledgements}

\bibliographystyle{aa}
\bibliography{referencias}

\end{document}